\documentclass[final]{svjour2}
\usepackage{graphicx}
\usepackage{rotating}
\usepackage{amssymb}
\usepackage{mathptmx}
\usepackage[numbers]{natbib}
\makeatletter
\journalname{Journal of Low Temperature Physics}

\bibpunct{}{}{,}{s}{}{,}

\begin{document}

\newcommand{\hdblarrow}{H\makebox[0.9ex][l]{$\downdownarrows$}-}
\newcommand{\um}{$\mathrm{\mu}$m\ }
\newcommand{\Lc}{$L_\mathrm{c}$\ }
\newcommand{\Lres}{$L_\mathrm{res}$\ }

\title{Integrated Filterbank for DESHIMA: A Submillimeter Imaging Spectrograph Based on Superconducting Resonators}

\author{A.~Endo$^1$ \and P.~van~der~Werf$^2$ \and R.M.J.~Janssen$^1$ \and P.J.~de~Visser$^{1,3}$  \and T.M.~Klapwijk$^1$ \and J.J.A.~Baselmans$^3$ \and L.~Ferrari$^4$ \and A.M.~Baryshev$^{4,5}$ \and S.J.C.~Yates$^3$}

\institute{1:Kavli Institute of NanoScience, Faculty of Applied Sciences, Delft University of Technology,\\ Lorentzweg 1, 2628 CJ Delft, the Netherlands\\
Tel.:+31 15 27 86113\\ Fax:+31 15 27 81413\\
\email{A.Endo@tudelft.nl}
\\2: Leiden Observatory, Leiden University, PO Box 9513, NL-2300 RA Leiden, the Netherlands
\\3: SRON, Sorbonnelaan 2, 3584 CA Utrecht, The Netherlands
\\4: SRON, Landleven 12, 9747 AD Groningen, the Netherlands
\\5: Kapteyn Astronomical Institute, University of Groningen, P.O. Box 800, 9700 AV Groningen, The Netherlands
}

\date{XX.XX.2011}

\maketitle


\begin{abstract}

An integrated filterbank (IFB) in combination with microwave kinetic inductance detectors (MKIDs), both based on superconducting resonators,
could be used to make broadband submillimeter imaging spectrographs that are compact and flexible. 
In order to investigate the possibility of adopting an IFB configuration for DESHIMA (\underline{De}lft \underline{S}RON \underline{Hi}gh-redshift \underline{Ma}pper), we study the basic properties of a coplanar-waveguide-based IFB using electromagnetic simulation. 
We show that a coupling efficiency greater than 1/2 can be achieved if transmission losses are negligible.
We arrive at a practical design for a 9 pixel $\times$ 920 color 3 dimensional imaging device that fits on a 4 inch wafer, 
which instantaneously covers multiple submillimeter telluric windows with a dispersion of $f/df = 1000$.

\end{abstract}

\section{Introduction}

\begin{figure}
\begin{center}
\includegraphics[%
  width=1\linewidth,
  keepaspectratio]{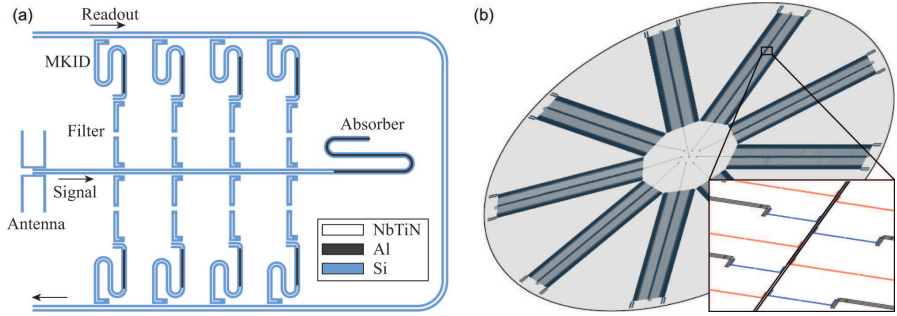}
\end{center}
\caption{(Color online) (a) Schematic of the integrated filterbank. The length scales are modified to enhance visibility. 
Each detection channel consists of a filter, made of a pair of lossless resonators which are resonant at a half wavelength of the signal ($\sim$100 \um), and an MKID, made of a lossy resonator which is resonant at quarter wavelength of the readout tone ($\sim$ 5 mm). The length of the resonators is made slightly different from channel to channel. The lossless section is made out fully of NbTiN (white), where the lossy section uses Al (black) in the center strip of the coplanar waveguide. The gray (blue in color) section indicates the region where there is no metal. At the end of the signal line is a matched absorber. (b) Example of a layout of DESHIMA on a 4-inch diameter wafer. There are 9 spiral antennas placed in the center with a nearest-neighbor pixel distance of 2 mm. From each antenna stretches a coplanar waveguide towards the integrated filterbank. The filterbank is designed to cover 320-475 GHz and 600-950 GHz with a dispersion of 1000 using 920 channels per pixel. The readout signal can access the MKIDs from the bonding pads located near the edge of the wafer. The inset is a close-up view of the filters of some channels. The longer filters (red) for the low-frequency band are spaced $\lambda /2$ away from each other, where the shorter filters (blue) for the higher frequency band are spaced $\lambda$ away from each other.
}
\label{cad}
\end{figure}

Submillimeter galaxies (SMGs) are massive star forming galaxies found in the early universe.
The distribution of SMGs across a wide range of redshifts $z$ is of great importance for studies of the cosmic history of star- and galaxy-formation, and the evolution of the cosmic large scale structure\cite{Amblard11}.
While SMGs are expected to be found in vast quantities in the coming years, 
by virtue of large format 2D submillimeter imaging cameras,
the measurement of the third dimension---redshift $z$ containing the history---would most likely become the rate-limiting step.
An ideal instrument for quickly measuring the $z$ of SMGs would be a submillimeter direct detection spectrometer with 
a dispersion $f/df\sim 1000$, to match the typical line width of SMGs with rotation velocities of $\sim$300 $\mathrm{km\ s^{-1}}$.
Although the first-generation of these so-called $z$-machines has successfully detected lines from high redshift sources\cite{Stacey11,Inami08},
the instantaneous bandwidth and number of pixels are limited by both the number of detectors and the optics. 
Moreover, the spatial sampling has been limited to 0 or 1 dimensions.

DESHIMA (\underline{De}lft \underline{S}RON \underline{Hi}gh-redshift \underline{Ma}pper) is a project to build a submillimeter imaging spectrograph using the advantage of multiplexability of microwave kinetic inductance detectors (MKIDs).
The number of detectors for MKID cameras\cite{Baryshev11} are quickly exceeding 10,000.
If these detectors are to be used for an imaging spectrograph, 
it is enough for tens of pixels to instantaneously cover
multiple submillimeter telluric windows with sufficient frequency resolution. 
Such an instrument would be ideal for submillimeter telescopes such as the 
Atacama Pathfinder EXperiment (APEX) and the Cornell Atacama Submillimeter Telescope (CCAT).
However, it is challenging to build such a spectrometer using grating optics, from the perspective of size, complexity,
and also to achieve a convenient spatial sampling.

One of the alternative frequency-selecting elements is a filterbank.
Filterbanks have been commonly used at microwave frequencies, and it is known\cite{Tauber91} that a pair of 
coupled coaxial cable resonators can be used as a band pass filter with an attractive dispersion of $~$1000.
At submillimeter wavelengths, such a filterbank will scale down in size to where the entire filterbank will fit onto 
a single chip, together with the MKIDs. Such an integrated filterbank (IFB) could take advantage of the compactness, 
flexibility and mass-production capability that superconducting thin-film technology offers also to MKIDs.

In this study, we investigate the possibility of adopting an IFB design for DESHIMA. 
We use numerical simulation to study the transmission of the IFB, and subsequently 
develop a layout in combination with MKIDs to be tested in laboratory and used on the sky.

\section{Numerical Simulation of the Integrated Filterbank}

\begin{figure}
\begin{center}
\includegraphics[%
  width=0.6\linewidth,
  keepaspectratio]{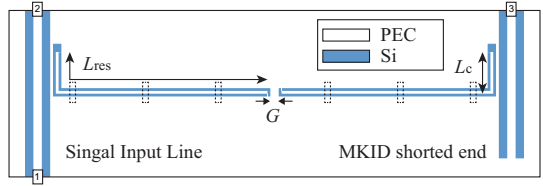}
\end{center}
\caption{(Color online) 3-port, 2.5 dimensional model used for the numerical simulation of the transmission of the filters.
The model consists of coplanar waveguides (CPWs) made of a perfect electric conductor (PEC, white) on 500 \um of Si (gray, blue in color), suspended in air.
The vertical CPW on the left with ports 1 and 2 represents the line from the antenna carrying the signal.
The vertical CPW on the right connected to port 3 represents the shorted end of the quarter wavelength microwave kinetic inductance detector (MKID).
The center strip and the slots of these two lines are all 3 \um wide.
In between the two lines is the filter, which consists of a pair of half wavelength resonators.
The CPW for the filter has a center strip and slots which are all 1 \um wide. The length of the entire resonator, the coupler, and the 
width of the ground plane between the two resonators are defined as $L_{\mathrm{res}}$, $L_{\mathrm{c}}$, and $G$, respectively. 
The dotted rectangles represent airbridges, also made of PEC.
}
\label{sonnet}
\end{figure}

The integrated filterbank design considered in this study is depicted in Fig.~\ref{cad}(a, b).
We select here a structure based on coplanar waveguides (CPWs) because 
lens-antenna coupled, NbTiN/Al hybrid MKIDs have demonstrated\cite{Yates11} photon noise 
limited sensitivity under loading powers as low as 100 fW: 
the single-channel loading power expected for a typical observation using DESHIMA. 
The signal shining on the antenna will transmit through a CPW line (signal line, hereafter) to which multiple filters are coupled.
Each channel of the filterbank consists of two sections; the first is a pair of lossless resonators which act as a narrow bandpass filter,
and the second is a lossy resonator which functions as an MKID. 
The MKID is a quarter wave resonator, coupled to the filter on its shorted end and to the readout CPW line on its open end. 
At the end of the signal line is a long, lossy CPW that absorbs the uncaptured signal to suppress standing waves.
The lossless section can be made of superconductors with gap frequencies higher than 1 THz, 
such as NbTiN or NbN. NbTiN is also preferable for the MKID section because of its low two level system noise\cite{Barends10}. In the lossy sections, the center strip of the CPW is replaced by a superconductor with a gap frequency lower than the signal, such as Al or TiN.

\begin{figure}
\begin{center}
\includegraphics[%
  width=0.8\linewidth,
  keepaspectratio]{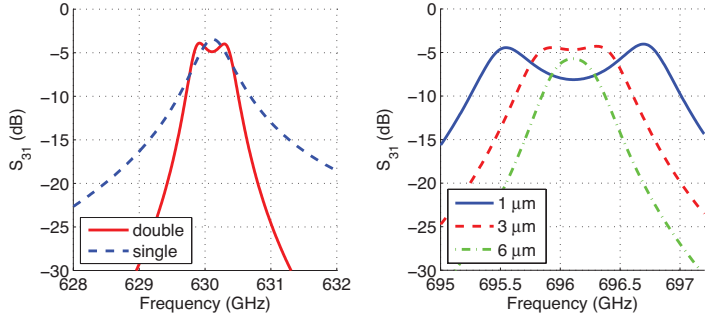}
\end{center}
\caption{(Color online) a) Calculated transmission $S_{31}$ of filters with 1 (dashed) and 2 (solid) halfwave resonators.
The geometry of the simulated model is described in Fig.~\ref{sonnet}, with $G=3$ \um and $L_{\mathrm{c}}=15 \ \mathrm{\mu m}$. 
\Lres has been adjusted so that the center frequency is equal in both cases. b) Transmission $S_{31}$ of a filter with 2 coupled resonators with varying gap width $G$.
}
\label{S31ab}
\end{figure}

The integrated filterbank is modeled using a commercial 2.5 dimensional electromagnetic simulator {\it Sonnet}\cite{Sonnet12}. 
We model the filterbank alone by replacing the antenna, the absorber,
and the connection from the shorted end to the rest of the MKID, by matched ports. 
The NbTiN film is replaced by a perfect electric conductor (PEC), assuming that the ohmic loss in NbTiN is negligible. 
The geometry of a single channel filter is shown in Fig.~\ref{sonnet}. 
As seen in Fig.~\ref{S31ab}~(a), the use of two coupled halfwave resonators adds a flat-top transmission and 
a sharper passband, compared to the case where a single resonator is used as a filter.
While the center frequency of the filter is determined mainly by the total length of the resonators, $L_{\mathrm{res}}$,
varying the inter-resonator ground plane width $G$ influences the frequency splitting between the symmetric- and anti-symmetric modes in the 
coupled resonators, as shown in Fig.~\ref{S31ab}~(b). 
The airbridges across the filters are required for suppressing the slotline mode,
which are observed to create an additional transmission peak when the bridges are absent. 
In practice, airbridges of Al can readily be made across NbTiN groundplanes\cite{Lankwarden11}.

Using the calculated scattering parameters of the single-channel transmission, we construct a 70 port network model of a 68 channel filterbank. 
In doing this we assume that the inter-coupling between 
each filter through free space and the substrate is negligible.
The geometry of each channel is described in Fig.~\ref{sonnet}, with $G=3$ \um and $L_{\mathrm{c}}=15 \ \mathrm{\mu m}$, 
and \Lres varying from 91.75 \um to 96.775 \um with a step of 75 nm.
After calculating the scattering parameters of individual channels, 
a network model is constructed by connecting the signal line ports of each channel 
with an ideal transmission line of half wavelength in between. 
Ports 1 and 2 of the network model are placed at each ends of the signal line,
and ports 3-70 are placed at the MKID-ends of the 68 filters.

\begin{figure}
\begin{center}
\includegraphics[%
  width=1\linewidth,
  keepaspectratio]{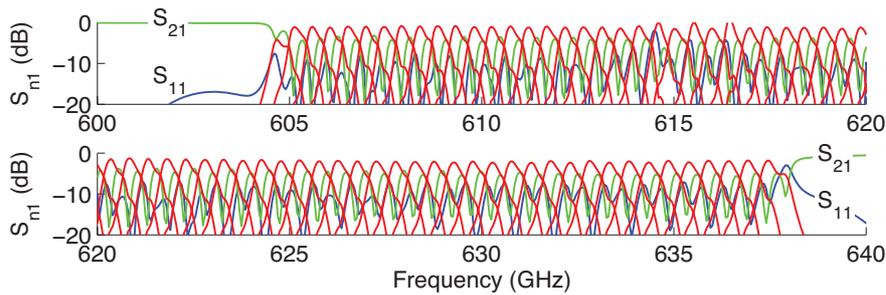}
\end{center}
\caption{(Color online) Calculated scattering parameters $S_{n1}$ for a 70-port model of a filterbank with 68 channels.
Curve $S_{11}$ represents the fractional power reflected back to the direction of the antenna.
Similarly, curve $S_{21}$ represents the power that passes all filters and flows into the absorbing termination.
The remaining curves are $S_{n1}$ ($n = 3,\ 2,...,\ 70$), which represent the power that flows into the individual channels to be absorbed by the microwave kinetic inductance detectors. 
}
\label{68bank}
\end{figure}

The scattering parameters of the complete network are plotted against frequency in Fig.~\ref{68bank}.
The amplitude of $S_{n1}$ representing the power flowing into the MKID is similar from channel to channel
across a bandwidth of 30 GHz, except for the few channels at the edge.
Strikingly, the transmission to the MKID has significantly increased from the case of a single isolated filter;
compare Fig.~\ref{S31ab}~(a) with Fig.~\ref{68bank}.
This can be understood as a result of multiple reflections between channels.
The unabsorbed fractional power ($S_{11}+S_{21}$) is less than $-3$ dB throughout the entire band, implying that more than 
half of the signal will be coupled to the MKIDs.
Note that the frequency spacing of the filters has been chosen rather arbitrarily
to simplify the simulation, 
and the spectral shape and inter-pixel overlapping of the transmission can be further optimized. 

Finally, leakage from the readout line to the signal line through the filter and the MKID at resonance has been calculated to be $<$$-50$ dB.

\section{Design of an Integrated DESHIMA chip}

Based on the simulated results, we construct a prototype design for an integrated DESHIMA chip.
In Fig.~\ref{cad}(b) we present an example of a 9 pixel $\times$ 920 color layout on a 4-inch wafer, 
which includes the antenna, the IFB, the MKIDs, the signal line, and the readout line.
In the center of the wafer are 9 CPW-fed log-periodic antennas, 
spaced 2 mm away from the nearest-neighboring pixels.
The number of channels is enough to instantaneously cover 320-475 GHz and 600-950 GHz with a dispersion of 1000.
In order to avoid the higher order mode of the low frequency channels from interfering with the high frequency channels, quarter wave resonators are used instead of the half wave resonators as filters.
In order to achieve a high channel density, while keeping the channels neighboring in frequency 
$n \times \lambda/2$ apart from each other, the high- and low-frequency bands are interwoven as shown in the inset of Fig.~\ref{cad}(b). 
The remaining empty space on the wafer is enough to double the number of pixels. 
The area per channel is limited mainly by the length of the MKID, 
which could be significantly reduced by replacing the distributed CPW resonators with parallel plate lumped element resonators\cite{Weber11}.
While our primary aim is to bring this IFB-based DESHIMA to ground-based submillimeter wave telescopes, 
the compactness and flexibility of the device compared to a grating-optics spectrometer 
makes it suitable for space- and air-born missions as well. 

Finally, we consider the issues that have not been taken into account in the simulation. Effects such as the transmission loss of the CPW and frequency shift due to crosstalk and/or fabrication errors need to be addressed experimentally. The difference in length of the filters of neighboring channels are $\sim$50 nm at 950 GHz, which is still much greater than the resolution of electron beam lithography. 

\section{Conclusion}

The result of electromagnetic simulation yields a straight-forward solution for integrating the entire filterbank and detector array of DESHIMA onto a single chip. The technology required for fabrication is largely in common with lens-antenna coupled NbTiN/Al hybrid MKIDs\cite{Lankwarden11}, which have shown photon-noise limited sensitivity down to optical loading powers expected for spectroscopy on ground. Though the concept needs further demonstration through experiments, which are under preparation, such an integrated filterbank configuration could yield a new generation of $z$-machines with full-3D spectral imaging capability.

\begin{acknowledgements}
We would like to thank E.F.C. Driessen for his inspiring comments. 
AE is financially supported by NWO (Veni grant 639.041.023) and JSPS 
Fellowship for Research Abroad. TMK likes to thank the W.M.Keck Institute for Space Sciences  for partial support for his stay at California Institute of Technology, while this manuscript was being written.

\end{acknowledgements}

\end{document}